\documentclass[manuscript,screen]{acmart}
\usepackage[utf8]{inputenc}

\setcopyright{none}
\renewcommand\footnotetextcopyrightpermission[1]{}

\newcommand{\red}[1]{\textcolor{black}{#1}}

\newcommand{\jrex}[1]{\textit{\textcolor{purple}{[jrex]: #1}}}
\newcommand{\alan}[1]{\textit{\textcolor{orange}{[alan]: #1}}}
\newcommand{\shir}[1]{\textit{\textcolor{blue}{[shir]: #1}}} 

\newcommand{\cut}[1]{}

\title{Compact Data Structures for Network Telemetry}

\author{Shir Landau Feibish}
\affiliation{
	\institution{The Open University of Israel}
 \country{Israel}
}

\author{Zaoxing Liu}
\affiliation{
	\institution{University of Maryland}
 \country{USA}
}

\author{Jennifer Rexford}
\affiliation{
	\institution{Princeton University}
 \country{USA}
}


\begin{document}

\begin{abstract}
Collecting and analyzing of network traffic data (\emph{network telemetry}) plays a critical role in managing modern networks.  Network administrators analyze their traffic to troubleshoot performance and reliability problems, and to detect and block cyberattacks.  However, conventional traffic-measurement techniques offer limited visibility into network conditions and rely on offline analysis.  Fortunately, network devices---such as switches and network interface cards---are increasingly programmable at the packet level, enabling flexible analysis of the traffic in place, as the packets fly by. However, to operate at high speed, these devices have limited memory and computational resources, leading to trade-offs between accuracy and overhead.  In response, an exciting research area emerged, bringing ideas from compact data structures and streaming algorithms to bear on important networking telemetry applications and the unique characteristics of high-speed network devices.  In this paper, we review the research on compact data structures for network telemetry and discuss promising directions for future research.
\end{abstract}

\maketitle
\pagestyle{plain} 

\section{Introduction}
Network administrators rely on traffic measurements to manage performance problems, flaky equipment, and cyberattacks in their networks.  For example, traffic measurements can reveal unusual levels of packet loss and delay, indicative of network congestion.  Similarly, traffic measurements can show a host receiving traffic from many different senders, suggestive of a distributed denial-of-service (DDoS) attack.
\emph{Network telemetry}---collecting and analyzing traffic measurements---enables network administrators to diagnose these problems, such as identifying the traffic responsible for congestion or pinpointing the senders participating in a denial-of-service attack.  Measurement data also help network administrators model the effects of proposed configuration changes to alleviate these problems, such as redirecting some traffic onto a different path or dropping packets from suspected attackers.

\subsection{Traditional Traffic Measurement}
Unfortunately, traditional high-speed network devices offer only limited visibility into the traffic, due to the overheads of collecting and exporting the measurement data for subsequent analysis. 
For example, each link may report statistics like utilization and packet loss, using the Simple Network Management Protocol (SNMP), but only on the timescale of minutes.  In addition, devices may report more detailed packet-level information (using technologies like Netflow~\cite{netflow}, sFlow~\cite{sflow}, and IPFIX~\cite{ipfix-old,ipfix}), but only for a small fraction, such as 1 in 5000, of the packets~\cite{sung2016robotron}.

These measurements provide a high-level summary of network conditions, but they do not offer the timely, fine-grained information needed to drive real-time management decisions. For example, SNMP data can show that a link suffers from persistent congestion, but not that a microburst disrupted performance for a few tens of milliseconds.  Similarly, Netflow data can identify the applications contributing the most traffic, but not performance statistics (like round-trip times or the prevalence of packet reordering) that look across multiple packets in the same flow.  In addition, sending measurement data to a collector for analysis introduces delay in reacting to changes in the network.  Fast reactions are important for alleviating congestion for real-time applications (such as video conferencing or self-driving cars) or blocking cyberattacks that are overwhelming victims. 

Perhaps most importantly, traditional measurement techniques cannot be \emph{customized} for the telemetry task at hand. SNMP and Netflow are useful for a variety of purposes, but they are not the best solution for any one telemetry task.  Sometimes, the total traffic volume on a one-minute timescale (as in SNMP) is the right statistic, but often it is not.  Similarly, sometimes packet samples of particular header fields (as in NetFlow) are the right data, but often they are not.  Network administrators need effective ways to customize the measurements they collect and the analysis they do.  They need effective ways to extract specific information from each packet, and combine that information across successive packets to have fine-grained visibility into network conditions at scale.

\subsection{Programmable Network Devices}


The emergence of programmable network devices, including switches and network interface cards (NICs), is poised to change all that.  
\red{Network devices have long run software in the \emph{control plane} to 
track changes to the network topology and compute routes.
Now,} the \emph{data plane} of network devices is increasingly programmable at the level of individual packets, with flexible parsing and computation based on packet header fields, as well as memory for accumulating information across successive packets. A prominent early example is the Reconfigurable Match Table (RMT) architecture~\cite{pisa} that was the basis for the Intel Tofino chipset~\cite{tofino}, which has been a popular platform in the networking research community. Other examples of programmable switches include Broadcom Trident~\cite{trident}, Juniper Trio~\cite{trio},  Aruba CX 10000~\cite{aruba},
and \red{Xsight X2~\cite{oxide}}. 
These switches are programmed using domain-specific languages like P4~\cite{p4,p416} and NPL~\cite{npl}.
In addition, programmable (\emph{smart}) NICs are available from various vendors, including Netronome~\cite{netronome}, Pensando~\cite{pensando}, Mellanox~\cite{mellanox}, and Xilinx~\cite{xilinx}.  Whereas high-speed switches often use programmable ASICs (application-specific integrated circuits),  SmartNICs often use technologies such as Field Programmable Gate Arrays (FPGAs) and multi-core engines that offer greater flexibility but necessarily operate at lower speed.

Flexible packet processing on network devices enables network administrators to customize the measurement data to the task at hand by collecting, analyzing, and even acting on the measurement results directly as the packets fly by.  For example, a network switch could identify the traffic responsible for a backlogged queue, and mark or drop the offending packets to alleviate the congestion.  As another example, the switch could identify the IP addresses of servers sending response traffic that does not correspond to any recent client request, and drop the unsolicited traffic.  As yet another example, the switch could identify senders contacting a large number of distinct receivers, or receivers contacted by a large number of distinct senders, to detect and mitigate denial-of-service attacks.  These and other telemetry tasks capitalize on programmability to group related packets with common header fields into a \emph{flow} and then compute and store per-flow statistics ranging from simple counts to more sophisticated performance and security metrics.

Unfortunately, high-speed network devices have significant resource limitations. These devices are domain-specific processors designed to process packets at high speed to keep up with link capacity. As such, these devices can only perform simple operations and maintain limited state.  Plus, since memory bandwidth is not keeping pace with link speed, the number of \emph{accesses} to memory for each packet is limited. In practice, many of these devices consist of a sequence of pipeline stages, each with match-action tables (for pattern matching on packet header fields), small register arrays (for storing data across successive packets), and simple arithmetic logic units (for performing addition, subtraction, and bit-wise operations).
Moreover, these devices must devote many of these resources to perform routine packet forwarding, leaving fewer resources available for traffic measurement.


\subsection{Compact Data Structures for Telemetry}
Luckily, many telemetry questions do not require exact answers; often a reasonable estimate is fine.  For example, identifying the most heavy flows may be more important than knowing the exact number of bytes or packets in these flows, let alone the sizes of the many smaller flows. Often network administrators care about a small fraction of the flows---the outliers---and a rough estimate of the associated statistics. Network administrators can exploit this tolerance by using \emph{approximate} data structures that trade accuracy for lower measurement overhead.  There is a long and rich history in the theoretical computer science community of research on compact data structures that can help design approximate solutions that ``fit'' in the data plane. However, past research on compact data structures does not always apply directly to high-speed programmable data planes. In particular, the constraints on the number of memory accesses, and the division of memory and processing across stages, do not arise in most earlier research on compact data structures.  

Over the past few years, we have seen great progress in designing compact data structures for high-speed network devices.  Many of these designs are variants of earlier compact data structures, tailored to the unique constraints of high-speed packet processing. 
Recent work shows how to support a wide range of important measurement tasks, both within a single network device and across a larger network.  In this paper, we present a survey of recent research on compact data structures for network telemetry, with an emphasis on how to grapple effectively with the unique constraints of modern network devices.
Our goal is to reach both the networking and the theory communities to foster further interdisciplinary collaborations in this area. The paper exposes theoretical computer scientists to a distinctive computational model for streaming algorithms, as well as a class of practical telemetry problems that need further study. For networking researchers, the paper puts a large body of recent research in a common context and shows how to adapt algorithms to the constraints of high-speed network devices.
\red{This is the first survey to focus on the data structures used for telemetry in programmable network hardware. Existing surveys on telemetry in the data plane have focused on other aspects of telemetry such as in-band telemetry that tags packets with measurement statistics~\cite{SurveyInBandTelemetry,SurveyPassiveInbandTelemetry}.  Broader surveys exist that look at the hardware and applications of programmable networks~\cite{SurveyProgDataPlane, SurveyStatefulDataPlane}, or the P4 language~\cite{SurveyP4ProgTaxonomy,SurveyP4ProgFundamentals}, that provide a wide-ranging discussion of telemetry in the data plane. Furthermore, surveys on sketches in general have also been presented~\cite{bloom_filters,SurveyCormode2011Sketch}. Our goal is to look at the overlap of these areas, in order to provide a much more extensive exploration of the data structures used for network telemetry to highlight both the challenges and considerations in creating and using these structures within programmable devices. }

The remainder of the paper is structured as follows. In Section~\ref{sec:telemetry}, we discuss the goals of network telemetry, and introduce a broad class of measurement tasks that perform queries on packet tuples. Then, Section~\ref{sec:dataplane} makes the case for supporting these tasks directly in the data plane, and introduces a computational model for high-speed network devices.  Section~\ref{sec:heavyhitters} starts our discussion of compact data structures for these devices, for simple queries that estimate set membership and per-flow traffic counts. Then, Section~\ref{sec:complex} considers more sophisticated queries for anomaly detection and performance monitoring.
Section~\ref{sec:practical} delves further into the practical challenges of allocating data-plane resources, to manage the trade-off between accuracy and overhead. 
The next two sections discuss early research in two promising directions: distributed network telemetry (Section~\ref{sec:distributed}) 
and robustness to adversaries
who try to manipulate the measurement process (Section~\ref{sec:security}).
 The paper concludes in Section~\ref{sec:conclude}.

\cut{
\begin{itemize}
\item Define flow (with flexible notion), telemetry
\item Distinction between management and control, and the role of telemetry in each
\item Picture of "before and after" of switches and control and management (and measure, analyze, act), and timescale
\item The shift from separation between measurement, analysis, and action, to integration of measurement and analysis, and even action
\item History of compact data structures in the TCS community and compare/contrast
\item Compact or approximate data structures (or sublinear), and at least two types (e.g., sketches and... space savings? cache? hash-table? key-value store?)
\item Traffic characteristics (number of flows, distributions of sizes)
\item Goal/audience for this paper
\item why are the constraints general? the exact form of the constraints will vary by target, but there are real constraints around moore's law, denard scaling, memory bandwidth, etc. intro?
\item Minlan's recent survey~\cite{telemetry-survey}
\end{itemize}
}

\section{Network Telemetry Queries}
\label{sec:telemetry}










Each packet in the network can be thought of as a \emph{tuple} of header fields (e.g., source IP address, destination port number, TCP sequence number, etc.)
and relevant attributes (e.g., packet size, timestamp, location, or queue length traversed).
A telemetry query runs over a stream of tuples by applying database-like operators inspired by platforms such as SQL or Map-Reduce~\cite{sonata,marple,newton,netqre}.

These queries often group related packets into a single \emph{flow}, such as a TCP connection or traffic between the source and destination IP prefixes. Precisely, a flow is a set of packets that share the same {\em flow identifier}, which is defined as packets sharing some tuple fields in common (e.g., the same $5$-tuple, which consists of the transport protocol, the source and destination IP addresses, and the source and destination port numbers). A flow telemetry query calculates one or more {\em metrics} based on the attribute associated with each flow and performs an aggregation on the packets of the flow, e.g., summation over packet count, bytes, or distinct number of flows. 

In addition, network administrators often perform telemetry queries on performance-related attributes, such as the packet timestamp. Estimating such statistics (e.g., round-trip latency) requires combining information across pairs of packets with stateful operations (e.g., the average round-trip time between a request packet and its acknowledgment).
The traffic measured in telemetry tasks is often defined by various time windows, or {\em epochs}, where an epoch represents a time period (e.g., several seconds to minutes). Table~\ref{tab:queries} summarizes common telemetry queries and their applications.



\subsection{Volumetric and Aggregated Flow Statistics}

\smallskip\noindent\textbf{Volumetric Flow Queries.} A common set of telemetry queries focus on the {\em size} of a flow, such as the number of packets  or the total byte count. With unlimited memory and compute resources, we could compute and \red{store} the sizes of all flows. However, network traffic is too large to be recorded at the per-flow level. Traditional telemetry approaches such as SNMP and NetFlow are usually too coarse-grained and report the measurements only on large epochs with sampled packets. Thus, the most popular queries in this class are $\alpha$-heavy hitters and Top-K flows. 
\red{
\begin{itemize}
    \item \underline{\em $\alpha$-Heavy hitters}: (also known as \emph{elephant} flows if measuring packet byte counts) are the large flows that consume more than a fraction $\alpha$ of the total traffic capacity. Network administrators can specify a fixed threshold $\alpha$ beforehand or set dynamic thresholds during the measurement.  
    \item  \underline{\em Top-$K$ flows}: are a variant of the heavy-hitter problem that reports the $K$ largest flows in the stream.  
\end{itemize}}

Volumetric flow queries are useful for a number of downstream network-management applications as heavy hitters are essential for network performance and security optimizations. For instance, traffic engineering~\cite{new_directions} needs to identify  the largest flows  and prioritize them to meet the service-level agreements. Similarly, in events of volumetric DDoS attacks~\cite{liu2021jaqen}, measuring heavy-hitter flows can guide further investigation. 


\smallskip
\noindent
\textbf{Aggregated flow statistics.} With the assumption of not recording the information for all flows, network operators also need to compute various aggregated statistical metrics that summarize all the flows. These metrics, such as entropy, quantiles, and distinct count, are concise representations of the flow size and metadata distributions, and are useful to monitor overall network conditions. 

\red{
\begin{itemize}
    \item \underline{\em Distinct count}: is the number of \emph{unique} elements seen in a stream. This can be a count of the number of different flows seen or the number of different attributes seen for a given flow. For example, for a given destination IP, we can measure the number of unique source IPs from which it is receiving traffic. A relatively high number of distinct source IPs could indicate that the destination is experiencing a DDoS attack. Moreover, the change in the number of distinct flows \red{(i.e., the number of unique flows in a stream of packets)} is a strong indicator for flash events (e.g., short-term traffic bursts) or ongoing DDoS attacks~\cite{liu2021jaqen,chen2020beaucoup,elasticsketch}. 
    \item  \underline{\em Quantiles}: is the general term for metrics such as median, $95^{th}$ or $99^{th}$ percentile. More broadly, quantiles can be inferred from the  cumulative distribution function (CDF). Quantiles are useful for identifying both short-term anomalies and and long-term changes in the traffic~\cite{qpipe}
    \item  \underline{\em Entropy}: is a measure of the diversity of the traffic~\cite{entropyCalculation} that can be used as an indicator for the distribution of the traffic. Anomaly detection may look into the distributions over different flow identifiers (e.g., source IP, destination IP) via continuously computing their entropy values~\cite{Entropy1} and identifying their changes~\cite{univmon}.
\end{itemize}}

\red{
Some of the metrics presented above are formally defined in Table~\ref{tab:queryDefinitions}. 
}

\begin{table}[h]
    \centering
    \begin{tabular}{l|p{5cm}|l}\hline
      \textbf{Metric}  & \textbf{Definition} & \textbf{Application} \\ \hline
         $\alpha$-Heavy hitters & All flows $f_i$ s.t. $|f_i| \geq \alpha \cdot n$ & Traffic Engineering~\cite{new_directions}\\\hline
         Top-$K$ flows & $\{f_1, f_2 ... f_k\}$& Load balancer~\cite{nitrosketch}\\\hline
        Distinct count (cardinality) & The size of the set ${\{f_1,f_2,...f_n\}:f_i\in S}$ & Attack detection~\cite{OpenSketch}\\\hline
        Entropy & $-\Sigma_{i=1}^j \frac{f_i}{n} log(\frac{f_i}{n})$ & Anomaly detection~\cite{univmon}\\
      \hline
    \end{tabular}
    \caption{\red{Flow statistics and queries. Given a stream of packets $S$, with $n$ packets  and $j$ distinct flows. Let $|f_i|$ denote the frequency of the $i$-th unique element in the stream s.t. $f_1 \geq f_2 \geq ... f_j$.}}
    \label{tab:queryDefinitions}
\end{table}



\subsection{Queries over packet metadata.} 

When packets traverse a network, multiple performance-related metrics can be recorded and attached to the packets as metadata, e.g., timestamps, switch locations, and queue lengths. These metadata can be used to compute various useful metrics about  network performance. For instance, by calculating the timestamp difference between outgoing and incoming traffic of a flow, we can estimate the round-trip time of the flow to a remote destination. 
\red{We can also identify if packets in a flow are seen out-of-order, which would normally cause retransmission and delays, possibly creating jitter and affecting the quality of experience of the network users. }
Moreover, by obtaining the queue occupancy information from each switch a packet traverses, we can understand the congestion status precisely and optimize the configurations for better performance accordingly.

\begin{table}[h]
    \centering
    \begin{tabular}{l|l|l}\hline
      \textbf{Attribute}  & \textbf{Metric} & \textbf{Application} \\ \hline
       Seq. No. & Out-of-order packets & Quality of experience (QoE)~\cite{lean}\\
      Timestamp & Round-trip times & QoE/Congestion~\cite{continuous_rtt}\\
      Timestamp & TTL changes & Network memory allocation~\cite{ttlMatters}\\ 
      Key & Set membership & Rate Limit~\cite{liu2021jaqen}\\
      Queue Length & Large queues  & Congestion Control~\cite{hpcc} \\
      \hline
    \end{tabular}
    \caption{\red{Example queries over packet metadata and applications.} 
    }
    \label{tab:queries}
\end{table}






\subsection{Queries over measurement windows. } Since network measurement data becomes less relevant with time, telemetry systems typically produce statistics about the most recently seen traffic.  For example, a network operator may want to know the top-ten flows over a $30$-second period.
With a \emph{tumbling} window, the time intervals do not overlap, and each packet is processed once and belongs to one window. For example, the first interval would correspond to times $0$ to $30$ seconds, while the next corresponds to times $30$ to $60$ seconds.
In contrast, a \emph{sliding} window slides over the stream of packets, always maintaining statistics for the most recent packets.  Sliding windows are more expensive to maintain, since the older packets expire gradually. In practice, a sliding window may be approximated using multiple smaller tumbling windows (e.g., times $0$-$10$, $10$-$20$, and $20$-$30$ for the $0$-$30$ second interval, followed by times $10$-$20$, $20$-$30$, and $30$-$40$ for the next interval covering times $10$-$40$).



\section{Case for Data-Plane Telemetry}
\label{sec:dataplane}
Although traditional measurement techniques have significant limitations, the emergence of programmable network devices enables unprecedented visibility into the underlying traffic.
Programmable data planes enable the \emph{customization} of telemetry at the packet level, for efficient fine-grained analysis and real-time adaptation to changing conditions.
However, in order to process packets at high speed, modern data planes impose significant limitations on memory and computational resources. Telemetry queries must ``fit'' within these resource constraints, as otherwise the data plane cannot serve traffic at line rate.



\subsection{Programmable Packet Processing}
Network devices, such as switches and network interface cards (NICs), are increasingly programmable at the packet level.  These  devices offer flexible: 

\begin{itemize}
    \item \emph{parsing} of packets, to extract specific fields of interest, 
    \item \emph{matching} on these fields to group related packets into a single ``flow,''
    \item \emph{computation}, such as simple hash functions as well as other arithmetic and logic operations,
    \item \emph{storage} of information across successive packets, and
    \item \emph{communication} with a software controller.
\end{itemize}


\noindent
Together, these capabilities make the data plane  a simple kind of \emph{stream processor} that performs operations over a sequence of packets. The programmable parser determines the ``tuple'' of fields extracted from each packet, the computation determines how to manipulate and update this data as the packet flies by, and the storage enables more sophisticated operations over multiple packets in the same flow. 
The data plane can generate reports to a software controller, and have the controller read the data-plane state and update the data-plane configuration.
These data planes are programmable using domain-specific languages such as P4~\cite{p4,p416} and NPL~\cite{npl}.


Programmable data planes are a promising way to enable customized telemetry, which offers several important advantages over traditional measurement techniques. First, performing both the measurement and analysis in the data plane improves \emph{efficiency}. The data plane can perform fine-grained analysis right as packets fly by, without exporting a large amount of data to the collector.
Second, the data plane can incorporate local \emph{metadata}, such as the current time or the current length of packet queues, into the analysis of the traffic. Third, the data plane can take timely \emph{action} on individual packets based on the results of the analysis.  For example, the data plane can drop, forward, or modify a packet in flight based on the results of the computation. 
Fourth, the data plane can protect \emph{privacy} by computing the answers to measurement questions without ever exporting the raw data used in the computation.

\subsection{Data-Plane Resource Constraints} 
\cut{
\begin{figure}
\centering
\includegraphics[trim = 0mm 0mm 0mm 0mm, clip,scale=0.4]{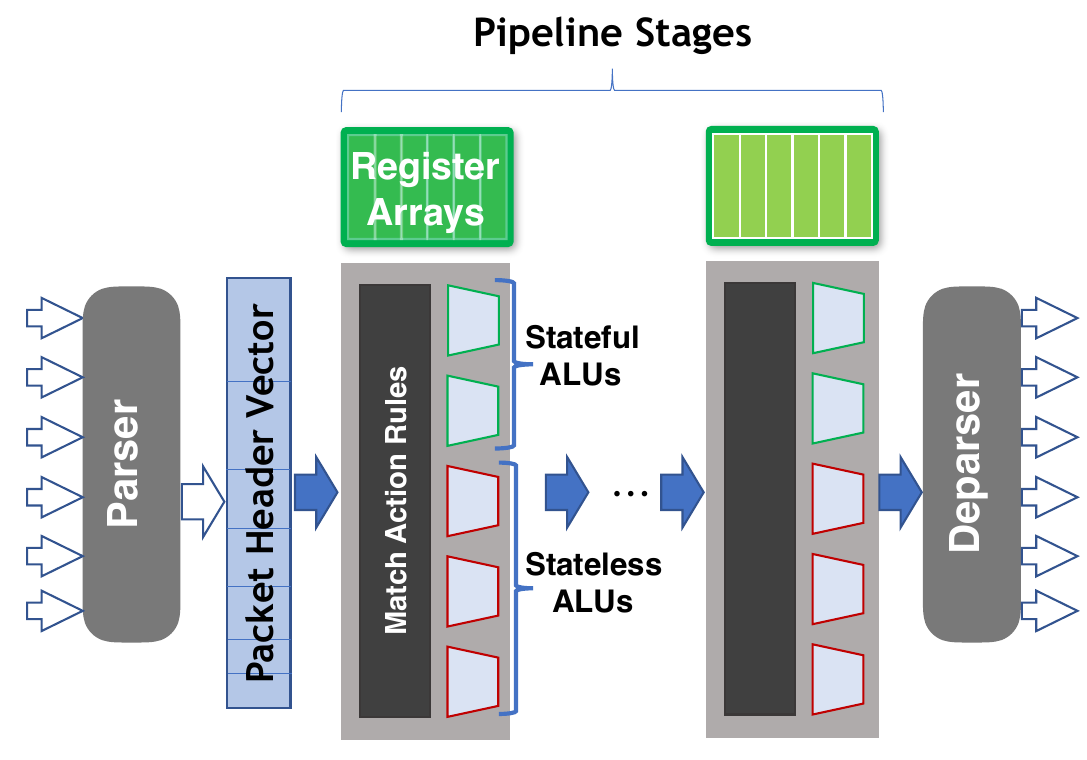}
\caption{PISA}
\label{fig:pisa}
\end{figure}
}


Unfortunately, programmable data planes have a number of limitations, due to the need to process packets at high speed. Typically, a high-speed data plane cannot parse arbitrarily deep into the packet, and computation is limited to simple arithmetic operations (e.g., addition and subtraction, but not multiplication and division) and logic operations (e.g., bit shifting). 
Since increases in memory bandwidth have not kept pace with link bandwidth, high-speed packet processing must work with limited memory resources. The memory is typically too small to store per-flow state. Plus, each packet can access the memory at most a small, constant number of times.

\begin{figure}
\centering
\includegraphics[trim = 0mm 0mm 0mm 0mm, clip,scale=0.6]{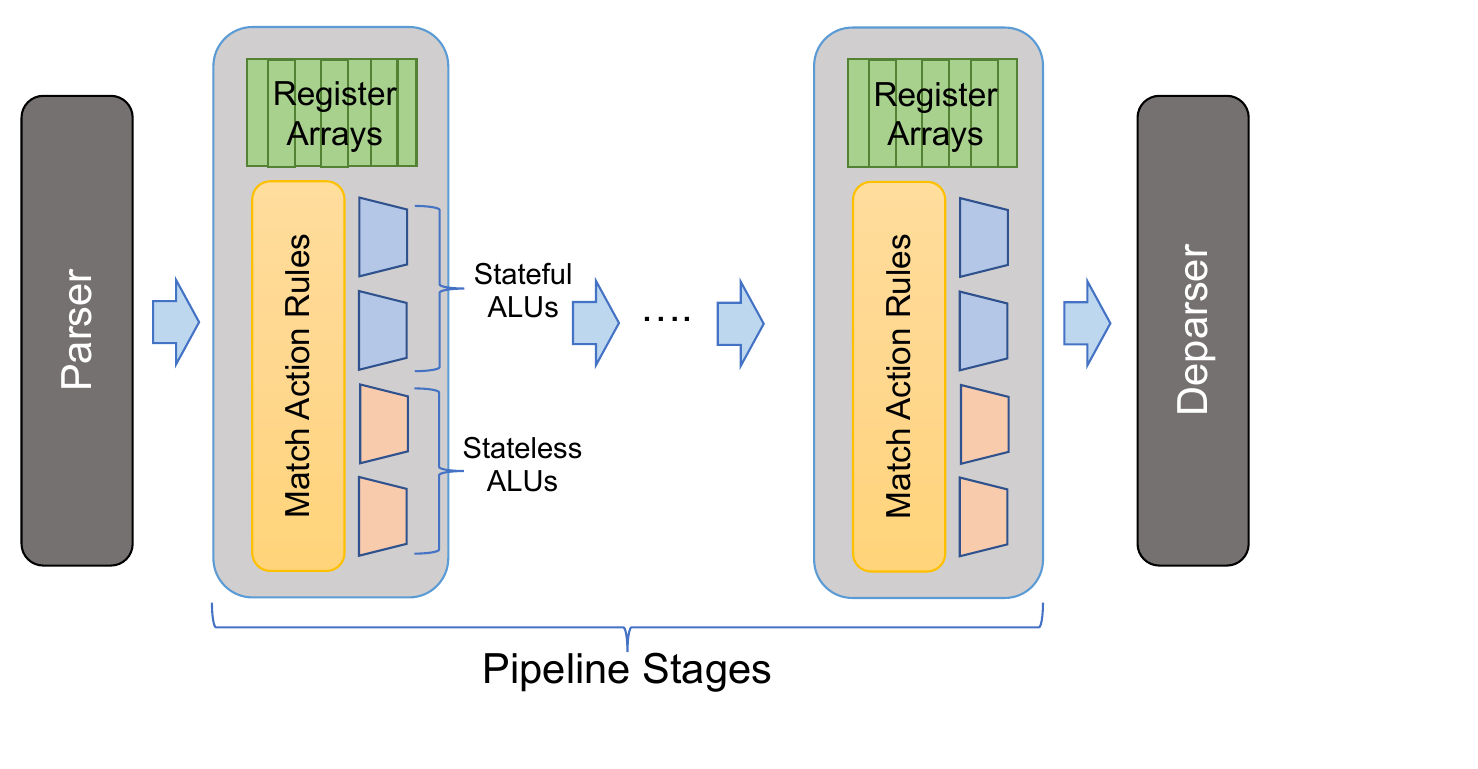}
\caption{Protocol-Independent Switch Architecture (PISA) Data Plane.}
\label{fig:pisaPipeline}
\end{figure}

The exact constraints differ from one kind of network device to another, but all of these devices have these kinds of limitations because high-speed links must be able to process a packet every few nanoseconds.  
A common computational model
is the Protocol-Independent Switch Architecture (PISA) where the data plane consists of a packet-processing pipeline with multiple stages of memory and processing resources.  As shown in Figure~\ref{fig:pisaPipeline}, each stage has:

\begin{itemize}
    \item small match-action tables that can perform exact or ternary \red{(i.e., matching to 0, 1, or a wildcard matching either)} matching of packet header fields based on rules installed by the control plane,
    \item simple arithmetic and logic units that perform actions, and
    \item small register arrays for storing information across successive packets,
\end{itemize}

\noindent
That is, the register memory is partitioned across the stages, where memory in an earlier stage cannot be updated based on the results of computations at a later stage, unless the packet is \emph{recirculated} to traverse the pipeline a second time.  
Each stage may have multiple parallel arithmetic/logic units and register arrays, allowing a single stage to perform multiple operations concurrently in the absence of dependencies. \red{Note that ALUs may be stateful or stateless. Stateful ALUs can perform actions on the register arrays in the memory of the stage. Stateless ALUs perform actions that do not access the registers in the stage, yet they may modify the packet metadata. }

\red{Figure~\ref{fig:pisaPipeline} presents a single pipeline. The pipeline is coupled to several physical ports in the switch. Furthermore, programmable switches normally have multiple pipelines such that each of the ports is coupled to a dedicated pipeline. An internal traffic manager forwards the packets between the pipelines. Packets may also be recirculated back to the begninning of a pipeline.  }
The data plane typically has limited bandwidth for recirculating packets, as well as limited bandwidth for communicating with the control-plane software.

Unfortunately, the resource constraints limit the accuracy of telemetry applications.   Fortunately, most telemetry applications are robust to small errors, enabling the use of approximate data structures that can work reasonably well with a limited amount of memory and a limited number of memory accesses per packet.  For example, a network administrator wanting to identify the heaviest flows may not mind if the estimates of per-flow traffic volumes have some errors. 

\section{Classic Data Structures} 
\label{sec:heavyhitters}






In this section, we give an overview of the data structures that are commonly used in network telemetry applications. Our goal is both
to acquaint networking practitioners with these classic data structures and to introduce theorists to the challenges of realizing these structures in the data plane. 


Many telemetry tasks rely on ``counting'' traffic volume (e.g., the number of bytes or packets) by flow, whether to estimate the count for each flow or to identify the heavy flows. Another common building block is ``set membership,'' where we need to represent a set of flows. 
To avoid maintaining per-flow state,
the data plane must store an approximate summary of the ``counts'' or the ``set.''

The approximation usually takes one of two forms: compressing all of the information (using a sketch) or discarding some information (using a cache).
These techniques have different characteristics, which can guide which data structure to apply to a particular setting. Cache-based approaches typically store the key for each cached entry, making the key easy to retrieve after the fact; sketches do not.  Sketches \red{are compact summaries of a data set that} can provide an approximate answer (e.g., an estimated count) on demand for any key; cache-based structures only produce estimates for the cached keys.
The data structures also differ in their accuracy, as well as whether they consistently overestimate or underestimate the statistic of interest.

Adapting these data structures to the constraints of high-speed data planes can be challenging. In the rest of this section, we first present the data structures that are the easiest to realize in the data plane, followed by those that require more substantial modifications. 
We first discuss two common sketches, the count-min sketch (for counting) and the Bloom filter (for set membership), followed by Space Saving that caches the large flows (for identifying the heavy-hitters).

\subsection{Sketch: Count-Min and Bloom Filter}

\subsubsection{Count-Min Sketch}
\begin{figure}
\centering
\includegraphics[trim = 0mm 125mm 125mm 0mm, clip,scale=0.48]{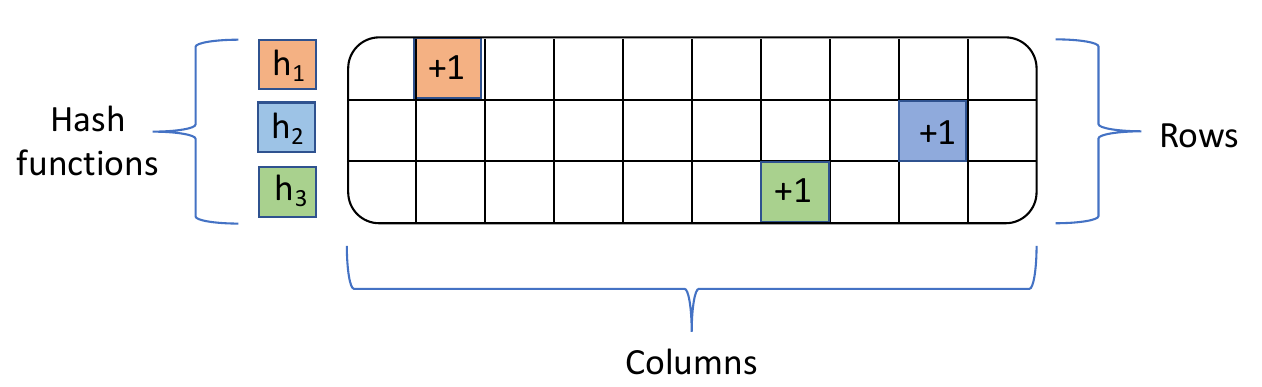}
\caption{Count-Min Sketch. \red{This is a sketch with $r=3$ rows and $c=10$ columns. Upon inserting a key it will be hashed by each of the hash functions and the relevant indices will be incremented by 1. } }
\label{fig:cms}
\end{figure}

To identify flows with counts that exceed a certain threshold, \red{various sketches have been proposed including the count sketch~\cite{CountSketch}, the count-min sketch~\cite{cms} and the CU sketch~\cite{new_directions}.} The count-min sketch~\cite{cms} is the de facto standard in programmable data planes, and is used extensively in telemetry applications~\cite{carpe,conquest,lean,liu2021jaqen,distcache,RHHH,netwarden}. 
As shown in Figure~\ref{fig:cms}, the count-min sketch is a matrix with $r$ rows and $c$ columns, where each of the $r*c$ entries stores a count. A set of $r$ independent hash functions is used to select an entry ($a_i$) in each row ($i$) for a given key. 
When updating the sketch, the associated counters of these indices are incremented. To estimate the count for a given key, the same hash functions are used to compute the same $r$ indices. The estimated count is the \emph{minimum} of the $r$ values.

The counters in the count-min sketch provide an approximate count for each key, yet this approximation may incur errors. Errors in the count-min sketch are due to collisions. Namely, if more than one flow maps to a certain index, the count is incremented when any one of these flows is seen, leading to over-estimation.  Note that the error is one-sided; that is, the counters may only \emph{over}-estimate the count but they never \emph{under}-estimate the count. 
Therefore, while the count-min sketch can answer count-queries for medium-sized flows and not just heavy flows, smaller flows may have higher error rates since they may be more significantly impacted by collisions with heavy flows.  
An interesting property of the count-min sketch is that it is linearly mergeable, meaning that several sketches can be combined into a single aggregated sketch. This property is especially useful for distributed telemetry, as discussed in Section~\ref{sec:distributed}.

\smallskip\noindent\textbf{Count-min sketches in the data plane. }
Generally speaking sketches fit very nicely within the constraints of the data plane. 
Each row of the sketch can be implemented as a register array within the memory of the switch. The number of columns is limited by the available memory. The number of hash functions, and thus the number of rows, which can be supported is limited by both the number of hash units available and the number of available memory accesses (since each row needs to be accessed exactly once when performing insert or get-count).

Count-min sketches are very space efficient structures. They do not require maintaining the keys of the flows, which could consume a lot of memory, and thus the size of the sketch remains constant, regardless of the size of the keys used. This is especially important in the data plane, since memory is statically allocated and dynamic allocation usually requires reconfiguring the switch. 
Nonetheless, the count-min sketch can only be queried if the key of the flow is known. 
If the key is not known, it cannot be hashed, and thus information about the flow cannot be retrieved. Therefore, the count-min sketch is often used to find the count of a flow as one of its packets is processed. 

\smallskip\noindent\textbf{Flow changes.}  The count-min sketch can also be used to identify flows that contribute the most to traffic change over two consecutive time windows. Consider two adjacent time windows $t_A$ and $t_B$. The size of a flow $i$ in $t_A$ is $S_A[i]$ and $S_B[i]$ in
$t_B$. The difference signal for $x$ is defined as $D[x] = |S_A[i]- S_B[i]|$. A flow is a {\em heavy} change flow if the difference in its signal exceeds the $\phi$ percentage of the total change over all flows. The total difference is
$D= \sum_{i\in[n]}D[i]$. A flow $i$ is defined to be a heavy change iff $D[i] \geq \phi\cdot D$. To detect such flow changes, we can take advantage of the intrinsic {\em linearity} in count-min sketch and count sketch~\cite{cms}, which measure two arbitrary time windows. When querying the flow changes, we need to perform a {\em subtraction} on each of the counters between two sketch instances and obtain the heavy hitters among the change in traffic volume. 

\cut{
While performing operations, such as insert or get-count, on a single flow in the data plane is straightforward, gathering general information about all flows from the sketch is more challenging. A count-min sketch can be queried only if the key of the flow is known. 
If the key is not known, it cannot be hashed and thus information about the flow cannot be retrieved. 
Therefore, while the control plane can copy the sketch it cannot effectively use it unless additional information is maintained. 
Information about each flow can be retrieved as the packet of the flow is being processed, though using this information is not trivial. 
For example,  we can check if the count of the flow has passed some threshold and report the flow. However, we may not want every packet of a flow to report, since that may overburden the communication with the controller. To overcome this, certain solutions use an additional structure to track flows~\cite{netseer} that have already been reported or use probabilistic reporting methods to lower the number of duplicate reports~\cite{carpe}. 
}

\begin{figure}
\centering
\includegraphics[trim = 0mm 100mm 125mm 0mm, clip,scale=0.48]{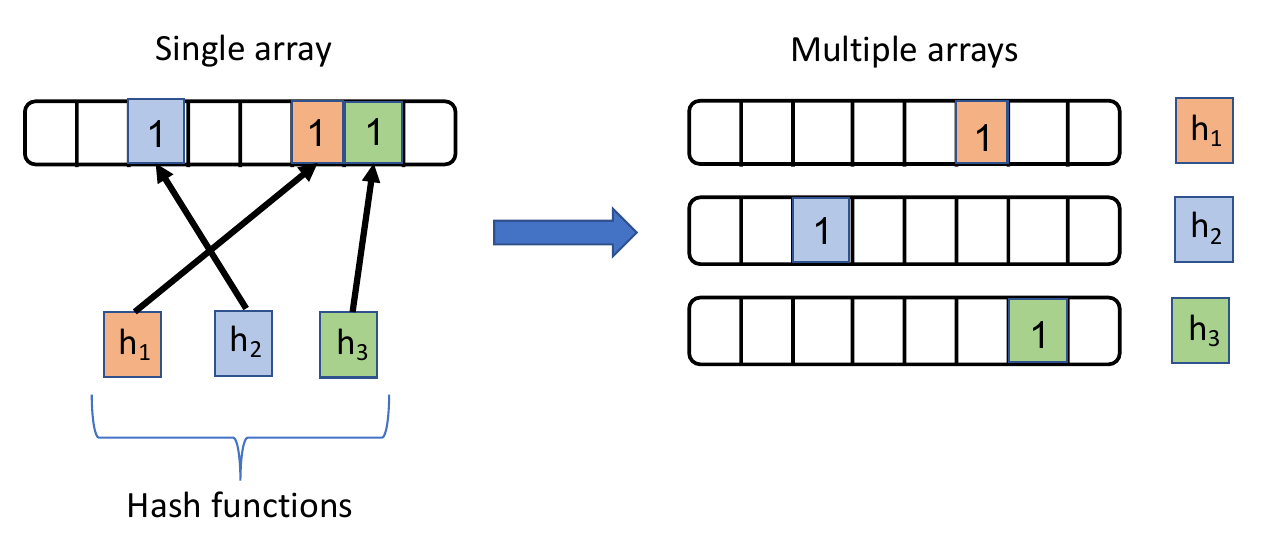}
\caption{Bloom filter in the data plane. \red{A bloom filter composed of a single array using three hash functions will be separated to three separate arrays in the data plane, and each hash function will be used to access one dedicated array.} }
\label{fig:bf}
\end{figure}

\subsubsection{Bloom Filter}
Another basic structure that is needed when processing traffic is a structure that maintains a set of flows, and supports set-membership queries. Due to the memory restrictions in the data plane, sets cannot be maintained in their entirety. Instead we can use a sketch called the Bloom filter~\cite{bloom_filters}. A Bloom filter is a sub-linear sketch composed of a single array consisting of $c$ bits, with $h$ independent hash functions that are associated with the structure. When processing an item, each of the $h$ hash functions is invoked on the key of the item to get $h$ different indices in the array. To insert an element into the structure, each of the associated bits is set to $1$. To perform a \emph{find} operation on an item, the same indices are checked. If all of them are set to $1$, the find operation returns true. Otherwise, it  returns false. A Bloom filter is always able to correctly identify items that have been inserted to the structure and  therefore does not have any false negatives. However, due to collisions, an item may appear to be in the set, even though it was never inserted to the structure; hence, false positives are possible.

\smallskip\noindent\textbf{Bloom filters in the data plane. }
While Bloom filters may seem to be a simpler structure (or just as simple) than a count-min sketch, surprisingly, implementing them in the data plane requires more adaptations. 
The main issue that arises is that both insert and find operations on a Bloom filters require setting or checking several indices in the array, and thus require \emph{multiple} accesses to the \emph{same} array. The memory model of the switch makes this impossible in a single iteration of packet processing.  In order to enable the use of multiple hash functions, the Bloom filter implementation in the data plane maintains a separate array for \emph{each} hash function. As shown in Figure~\ref{fig:bf}, for a Bloom filter with $c$ bits using $h=3$ hash functions, three separate arrays of size $c$ bits need to be maintained. This results in a slightly different implementation than the standard Bloom filter. However, since it avoids collisions between the different hash functions, it potentially has fewer collisions. Thus providing an error rate that is no higher than the error rate provided by the regular implementation of Bloom filter. 
As with the count-min sketch, the parameters of the Bloom filter are constrained by the available resources. 

\cut{
\shir{maybe remove resource constraints here.} Resource constraints affect the parameters of the bloom filter. The size of the array is limited by the available memory and  the number of hash functions is limited by both the number of hash units available and the number of available memory accesses.
Additionally, if the key is not known, it cannot be hashed and thus information about the flows cannot be retrieved. 
Therefore, while the control plane can copy the data structure it cannot effectively use it unless additional information is maintained.
}

\subsection{Cache: Space Saving}
One of the most basic structures is the key-value store. It is essentially a fixed-size hash table indexed by hashing a key, such that it effectively becomes a cache in a setting with limited memory, which can be used for both flow counts and set-membership. 
Upon insertion, an item is hashed to one index in the hash table. If the entry is empty, the item is inserted. Otherwise, the item may evict an existing item or may be discarded. In either case, an item from the set would not appear or be counted in the structure, and thus may create false-negatives in set-membership or under-estimation in flow count approximations. 
Hash tables may be implemented as a single table or as a table divided across multiple stages to enable better handling of collisions, though multi-stage tables do not eliminate the problem. 

\begin{figure}
\centering
\includegraphics[trim = 3mm 40mm 20mm 0mm, clip,scale=0.48]{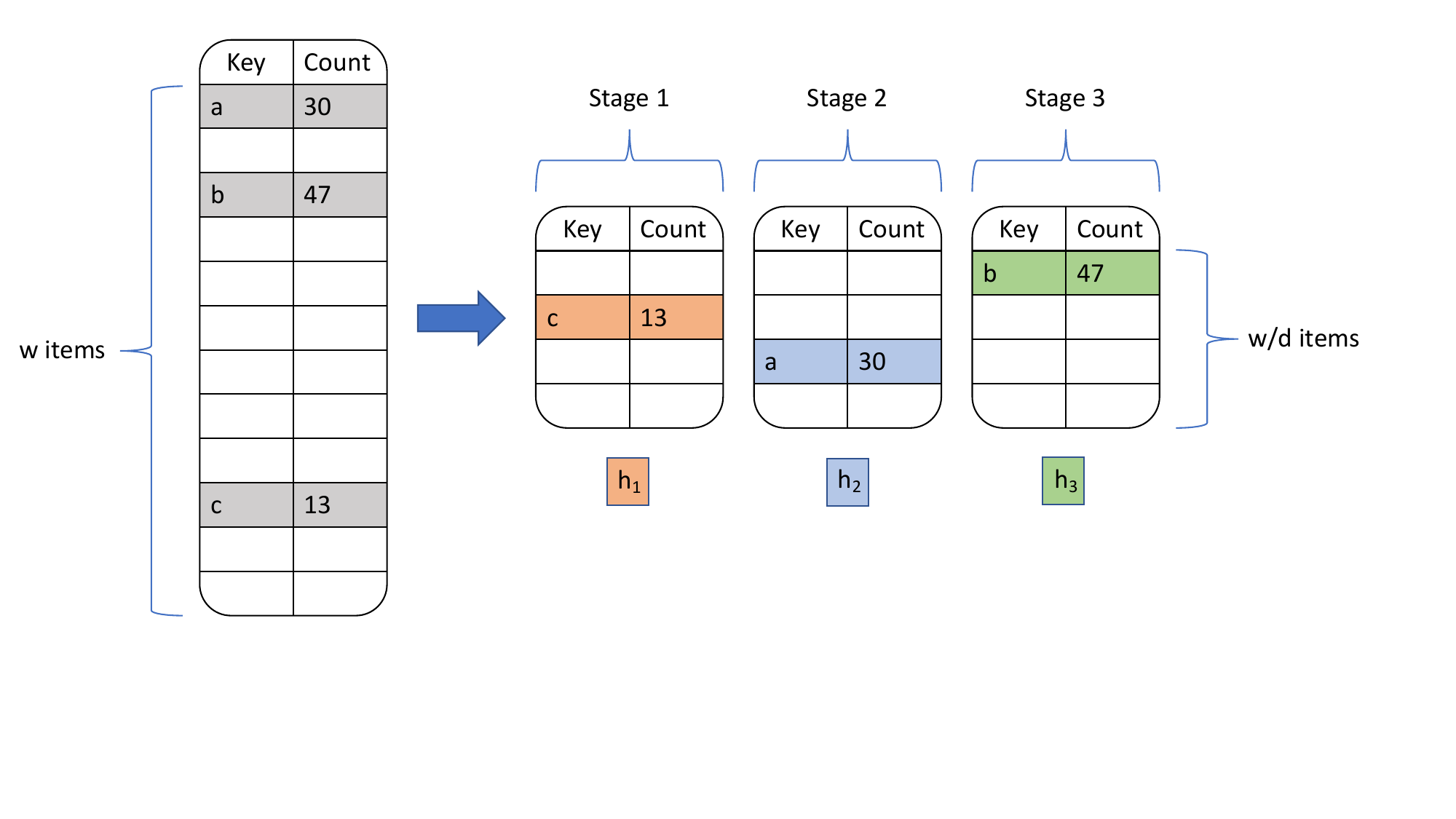}
\caption{Space Saving in the data plane. \red{The Space Saving table of size $w$ is divided into $w/d$ tables for $d=3$, with one part of the table in each stage. Items are hashed into each of the tables and thus may appear in any of $d$ tables.  }}
\label{fig:ss_prec}
\end{figure}

\noindent\textbf{Space Saving. }
There are several variants for the flow count problem, such as finding the heavy-hitter flows or the top-k flows. 
Several cache-based (also known as counter-based) algorithms exist in the theory literature for solving the heavy-hitter problem. 
Perhaps, the most widely used algorithm is the Space Saving algorithm of Metwally et. al.~\cite{SpaceSavings}. Space Saving maintains a table of size $w$, and works as follows: upon insertion of an item $x$, if $x$ is in the table, increment its counter by $1$. Otherwise, find the item with the smallest counter in the table and replace the key with $x$. The counter is kept (i.e., it is not reset) and incremented by $1$. To look-up the value of item $y$, traverse the table to find $y$ and output it's count. If $y$ is not found in the structure, \red{its frequency is at most the minimum counter in the table. Furthermore, if the frequency of any item in the stream is larger than the minimum counter in the table, it must exist in the table~\cite{SpaceSavings}. }

\noindent\textbf{Space Saving in the data plane.}
Performing this exact algorithm in the data plane is very problematic. Due to the limited number of memory accesses that may be performed while processing a packet, traversing the entire table would require numerous re-circulations and therefore would not be practical. 
This means that both finding an item $x$ in the table and looking for the minimum counter would not be possible. 

Several attempts have been made to adapt this algorithm to the data plane~\cite{HashPipe,precisionToN}. 
HashPipe~\cite{HashPipe} was the first algorithm to adapt the Space Saving algorithm to the data plane. Precision~\cite{precisionToN} later improved the performance of the algorithm by introducing probabilistic recirculation. Precision is based on a variant of Space Saving called RAP (Random Admission Policy)~\cite{RAP}.
As shown in Figure~\ref{fig:ss_prec}, the single key-value table of size $w$ is divided into $d$ hash tables of size $w/d$ each placed in a separate stage.
Precision works as follows: 
When inserting an item $x$, instead of traversing the entire table to find an item, it will be hashed using $d$ independent hash functions to one index in each of the $d$ tables. If $x$ is found in one of these indexes, it's counter will be incremented by $1$.
If an item is not found in the structure, it will be inserted with some \emph{probability} $p$. Therefore,  not every item seen will necessarily be inserted to the structure. To insert the item, it is recirculated and processed by the pipeline a second time. The probability for insertion (and recirculation) is based on the value of the minimum counter seen in the $d$ indexes, and decreases as the minimum counter increases. 
While Space Saving sketch may only overestimate the counters, the probabilistic recirculation in Precision introduces a two-sided error which may either over or under estimate the values of the counters.

\section{Complex Data Structures}
\label{sec:complex}


In this section, we discuss approximate data structures that support more sophisticated telemetry queries beyond estimating traffic counts for heavy flows. As described in the previous section, it is possible to make relatively straightforward adjustments to classic data structures such as the count-min sketch so they can function within the 
the data plane. For more sophisticated queries, traditional data structures may perform computations or memory accesses in complex ways that do not have a natural analogue in the data plane; instead, new designs are needed. So far, a wide variety of data structures for the data plane have been introduced to (1) estimate various flow-level statistics over a single flow key definition, such as distinct flows, entropy, and flow changes, (2) answer multiple queries over multiple keys, and (3) compute network performance statistics.




\subsection{Sketches for Distinct Counting}




Estimating the number of distinct flows
is a fundamental problem in network telemetry. Given a definition of a {\em flow key} (such as a source IP address or 5-tuple), the number of distinct flows 
is defined as the number of distinct keys appearing in the traffic. To estimate the number of distinct flows, we can consider the classic Linear Counting algorithm~\cite{whang1990linear} as an example. Linear Counting (LC) has simple data-plane logic to achieve fast packet processing~\cite{OpenSketch,elasticsketch} but cannot maintain high accuracy when the distinct count is large and the memory space is relatively small. At a high level, the LC algorithm needs to maintain a vector of length $m$ and uses a uniformly random hash function to map a flow into an index in the vector (i.e., an $m$-bit hash table). \red{When all flows are mapped to the hash table,} the key idea of LC is to leverage the number of ``collisions'' happening  among $m$ bits, which is an indicator of how many distinct flows are added into the data structure.
For example, assuming there are $n$ flows, we consider three possible cases: (1) If $n \ll m$, then the number of bits set to $1$ is a good approximation of $n$. \red{(2) If $n \approx m$, we need to check how ``full'' the vector is by counting how many bits are still set to 0. The number of 0s, $m_0$, can be used to estimate distinct flows as $\widehat{dist} = m \cdot ln(\frac{m}{m_0})$}. (3) If $n >> m$, the approximation will not work well, which is the fundamental limitation for this type of algorithm.

In practice, the LC algorithm is often used in conjunction with the count-min sketch to estimate the number of distinct flows as the LC algorithm is essentially the same as maintaining a row of counters~\cite{elasticsketch}. However, due to the inaccuracy of the LC algorithm in practice when the number of distinct flows is large (e.g., $m_0$, as the number of ``0'' counters, in the sketch becomes small or even $0$ as shown in~\cite{nitrosketch}), more accurate distinct counting sketches are needed. Solutions such as LogLog~\cite{loglog} and HyperLogLog~\cite{hyperloglog} (an extension to the \red{Probabilistic Counting algorithm of Flajolet–Martin~\cite{pcsa}}) are proposed to optimize memory efficiency of counting distinct elements. However, it is challenging to adopt them in the data plane due to a potentially large number of memory accesses. For instance, when adding a new flow into HyperLoglog, we need to perform complex operations on the hash output, such as using only the leftmost subset of the hash index as the counter address and finding the left most ``1'' in the remaining bits in the hash. Such complex operations manipulating the hash bits are challenging in current programmable data-plane hardware.

\smallskip\noindent\textbf{Count distinct above the threshold in the data plane:}  To control the number of memory accesses and avoid complex hash operations in estimating \emph{the number} of distinct flows, an alternative approach is to estimate whether the distinct count \emph{exceeds a given threshold} (e.g., whether the number of distinct flows is greater than 130). One such method is based on the  {\em coupon collector problem}. At a high level, the coupon-collector problem asks how many random draws (with replacement) are needed to collect all coupons at least once. For instance, we need 129.9 draws in expectation to collect each of 32 coupons. 
We therefore can use a 32-coupon collector to identify if the number of distinct flows is at least 130. In a recent effort, BeauCoup~\cite{chen2020beaucoup} explores this idea to build a distinct counting system in the data plane and shows that a coupon drawing process can be implemented efficiently in hardware with one memory access per packet. \red{BeauCoup uses a variant of the coupon-collection scheme that does not assume a uniform probability for collecting all coupons.  Instead BeauCoup sets a different probability for drawing each coupon to achieve the desired threshold with a limited number of coupons. }

\subsection{Sketches for Entropy}



In addition to estimating individual flow information, a recent focus has been on estimating metrics that represent entire flow distributions. We call these statistics  \emph{distribution measurements}.
Compared to estimating individual flows, a distribution measurement requires appropriate metrics to capture and summarize flow attributes
of the underlying traffic distribution. Standard statistics in measuring distributions are moments (e.g., standard deviation, mean, kurtosis for tailedness of a distribution, and skewness for the distortion of a distribution), but a more empirically useful statistic for networks is {\em entropy}, a succinct means of summarizing traffic distributions for anomaly detection~\cite{Entropy1}, DDoS attack detection~\cite{liu2021jaqen}, and fine-grained traffic classification~\cite{Chakrabarti_entropy}. For instance, a network administrator can track the entropy changes among multiple header fields (e.g., source IP, destination IP, and port number) to identify potential traffic anomalies. A widely adopted definition of entropy is the {\em Shannon entropy}, defined as $-\sum_{i=1}^n \frac{f_i}{m} \log(\frac{f_i}{m})$, where there are $n$ flows of total size $m$ and each flow $i$ has size $f_i$. 

Lall et al.~\cite{Entropy1} proposed an entropy estimation algorithm based on the idea of the celebrated AMS sketch~\cite{ams}. Conceptually, the algorithm can be divided into three steps. In the first step (random selection), the algorithm is prepared to select random locations in the packet stream. These locations decide the set of packets (their flow keys) that the algorithm tracks online.
Then in the second step (flow tracking), the algorithm will keep track of the number of packets in particular flows
since the selected packet orders/locations in the stream (i.e., determine which flows are tracked). Finally, in the third step, the entropy can be estimated through tracked flow counters and an offline estimation procedure via logarithmic operations and floating-point number calculations. In summary, this type of sketch algorithm needs to select a randomly chosen set of flows and maintain their flow sizes as the presentation of the entire flow size distribution.

\smallskip\noindent\textbf{Entropy sketches in the data plane.} It is often challenging to implement existing entropy sketches entirely in the data plane. Using the above entropy sketch as an example, its first and second steps can be implemented in the data plane, but the third step cannot. The first step is to randomly select a set of locations in a stream to start tracking flows. Instead of performing uniform sampling on each packet, the control plane needs to pre-compute random samples of locations in the upcoming packet stream and deploy these samples into the programmable data plane to check when packets fly by. Based on these random samples as the locations to start updating the sketch with associated flows, the data plane needs to perform random counter updates as the second step and augment the counters to compute entropy values. However, due to parallel memory accesses and complex entropy estimation operations required in the third step, it cannot be performed 
in the data plane. Other entropy estimation algorithms, 
such as~\cite{simple_entropy} and~\cite{harvey2008sketching_entropy}, are also infeasible in the data plane. Therefore, a pragmatic solution is to offload the final step of the entropy calculation to the control plane by reporting the sketch counters to the CPU~\cite{liu2021jaqen}. However,  retrieving sketch counters can incur non-trivial delays, due to the inefficient implementation of the control-plane API~\cite{namkung2021telemetry}. A batch-based implementation can speed up counter reading~\cite{namkung2021telemetry}. To further reduce the compute footprint and optimize the latency of such data plane and control plane communication, SketchLib~\cite{sketchlib} proposes several techniques to reuse hash and counter arrays to reduce the counters for entropy calculation. With recent efforts to support floating-point number calculations in the data plane~\cite{yuan2022unlocking,cui2021netfc, statsinP4, inRec}, it is a promising direction to design and implement entropy sketches entirely in the data plane to track real-time traffic changes. 





\subsection{Multi-Metric Sketches}

In practice, applications today require obtaining traffic metrics based on the keys defined by different tuple fields and their attributes. For example, traffic engineering \cite{traffic-engineer} in the host-level may use the source IP as a key to track heavy hitters, while flow scheduling \cite{flow-schedule} may need 5-tuple \red{(i.e., protocol, source IP, source port, destination IP and destination port)} as the key.
By providing application-specific key definitions, network administrators often need to perform multiple flow-level queries on the {\em same flow key} or the same query on {\em distinct flow keys}. For instance, administrators want to query heavy hitters, entropy, and distinct flows over 5-tuple flows using a sketch. Alternatively, for security detection and diagnosis, it is required to query metrics over multiple flow keys and sometimes it is even challenging to predict what specific keys are useful unless we exhaustively track all possible keys. 
Motivated by these use cases, sketch-based solutions have evolved that can \emph{simultaneously} query \emph{multiple} types of metrics and keys. We categorize these solutions into three categories:  ``multi-metric, same key'', ``single-metric, hierarchical keys'', and ``single-metric, separate keys''.

\begin{figure}[t]
\centering
\includegraphics[scale=0.6]{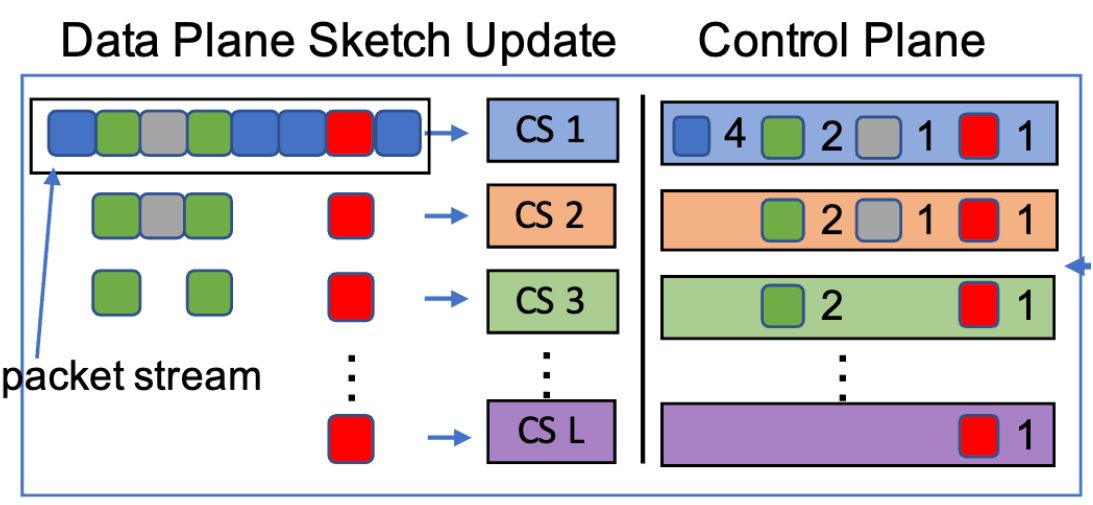}
\caption{UnivMon sketch overview.}
\label{fig:univmon}
\end{figure}

\smallskip\noindent\textbf{Multiple metrics on the same key.} 
The theoretical foundation of designing a single sketch to estimate multiple statistics comes from the concept of \textit{universal streaming}. The main question that universal streaming seeks to answer is whether such algorithms can be extended to estimate more general metrics of the form $g(f_i)$ for an arbitrary function $g$ defined over the flow distribution. We refer to this statistic as $G$-$sum$~\cite{zero_one_law}.
Conceptually, sketches that can estimate multiple statistics on the same key often consist of multiple single sketches looking at different subsets of the traffic to enable richer queries over the entire traffic. For example, UnivMon~\cite{univmon} uses \red{multiple instances of Count Sketches, called {\em levels}, to support the estimation of a wide range of traffic statistics, including heavy hitters, distinct flows, entropy, and flow changes.} \red{Based on its analysis, the number of levels needed is logarithmically depending the number of distinct flows in the traffic}.  At a high level, UnivMon leverages theoretical advances in {\em universal sketching}~\cite{universal_streaming_norms,zero_one_law}. When updating the sketch for each packet, UnivMon performs up to $L=O(\log n)$ hashes on the flow key (of the packet) that output a single bit $0$ or $1$, and starting from the first hash, it uses the longest sequence of $1$s to determine if a flow should be tracked at one or multiple levels, as shown in Figure~\ref{fig:univmon}. For example, if the first three hashes of a flow key are all $1$ but the fourth hash output is $0$, the hashing process stops and the first three Count Sketch instances will be updated with this flow.  
This construction of multiple levels of independent sketches is to ensure that the algorithm is able to capture a broad representation of flows for the size estimation across all flows. However, such a packet insertion process can take up to $L$ updates to all the $L$ levels of sketches. Recent efforts of SketchLib~\cite{sketchlib} and Sketchovosky~\cite{sketchovsky} proposed an efficient insertion operation by only updating the \red{relevant components of the sketch. }

 Other sketches such as FCM-Sketch~\cite{song2020fcm}, PCSA~\cite{pcsa}, MRAC~\cite{kumar2004data}, and multi-resolution bitmap (MRB)~\cite{multiresolution_bitmap} use multiple single arrays of sketch counters. To efficiently implement these multi-query sketches on programmable switches, recently SketchLib~\cite{sketchlib} has provided a comprehensive library to support the above-mentioned sketches. Recent work, Panakos~\cite{zhao2023panakos}, proposed a sketch that can estimate entropy, distinct count, and tail quantiles accurately, and it is implemented in a P4-based programmable switch. \red{To further improve the accuracy-memory tradeoff, BitSense~\cite{ding2023bitsense} proposed a compressive sensing framework to compress and recover counters in a sketch.}


 \smallskip\noindent\textbf{Single metric over hierarchical keys.} A representative traffic metric over hierarchical keys is hierarchical heavy hitters (HHH), where the hierarchy is determined based on the type of prefixes of interest in a given application. \red{The definition of hierarchical keys is motivated by the observation that IP addresses are naturally hierarchical. Suppose that a single entity controls all IP addresses of subnet $66.249.72.*$, where $*$ is a wildcard byte. It is possible for the controlling entity to spread out traffic uniformly among this set of IP addresses, so that no single IP address (e.g., $66.249.72.1$ and $66.249.72.2$) within the set of addresses $66.249.72.*$ is a heavy hitter, but we may want to know if the sum of the traffic of all IP addresses in the subnet exceeds a specified threshold as a heavy hitter.}
 For instance, anomaly and DDoS detection applications often require identifying frequent flow aggregates based on common IP prefixes, where each device may only generate a small portion of the traffic, but their combined traffic volume is overwhelming.

 To answer heavy hitter queries over hierarchical keys, a basic sketch construction for HHH detection is to run an independent sketch instance that detects heavy hitters per layer of the hierarchy. For instance, Random-HHH~\cite{RHHH} is a sketch that maintains a count-min sketch to track the heavy hitters per layer of the hierarchy. In this way, one can find all possible heavy hitters of all layers (e.g., all prefixes in IP addresses) and thus determine the hierarchical heavy hitters by aggregation. However, updating all sketch instances on all layers is computationally expensive  because there are potentially a large number of layers (e.g., 32 in IPv4). Instead of updating all layers, RHHH randomly selects one level of sketch instances using a level-specific key (e.g., IP prefix) to update per packet. The analysis of RHHH demonstrates that RHHH can achieve similar accuracy as the update-all construction when receiving sufficient packets.

In data plane hardware, maintaining multiple sketch instances for tracking HHHs is resource-heavy and often infeasible when the number of layers is large.
CocoSketch~\cite{cocosketch} tackles this problem by maintaining a single sketch instance and leveraging the hardware resources more efficiently. CocoSketch is designed to support heavy hitter queries over arbitrary flow keys in a pre-defined hierarchy. It is motivated by the challenge to predict what specific keys are relevant before the fact (e.g., security events and performance anomalies). The key insight behind CocoSketch shares the same theoretical basis with Unbiased Space Saving~\cite{ting2018data} to address the subset sum problem~\cite{priority-sample}. Given a set of items, each with a weight, the subset sum estimation problem estimates the total weight of any subset of items. The problem of arbitrary partial key queries can be cast as the subset sum estimation problem: the size of a partial-key flow $e$ equals the total size of a subset of full-key flows that match on the partial key with $e$. For instance, the size of a flow $e$ defined by the fields of source IP and destination IP equals the total size of all 5-tuple flows that share the source IP and destination IP with $e$. \red{Unbiased Space Saving is very similar to the original Space-Saving algorithm, yet in the case that the item is not in the table, instead of always replacing the item with the minimum counter, it will only replace it with a probability based on the value of the counter. Using the }  
Unbiased Space Saving technique, CocoSketch minimizes the variance of its subset-sum estimation for querying arbitrary flow keys.

 \smallskip\noindent\textbf{Single metric over separate keys.} A straightforward way to estimate a metric over different flow keys would require instantiating multiple separate data structures (e.g., using three count-min sketches to detect the heavy flows in source IP, destination IP, and 5-tuple). Having separate data structures would consume significant  memory space in the data plane. What's worse is that to maintain line rate, programmable switches only allow a small constant number of memory accesses per packet, making it infeasible to update multiple data structures for every packet.
 
 A line of recent work~\cite{chen2020beaucoup,sketchovsky,cocosketch} has focused on designing a single data structure to estimate a statistic/metric over separate flow keys. BeauCoup adopts the idea of the coupon collector problem to estimate if the number of distinct key values is larger than a threshold for many separate keys (e.g., any packet header fields). For example, BeauCoup can be used to estimate if the number of distinct destination IPs from a source IP is above a certain threshold, which is considered as a ``superspreader''. Interestingly, with the ability to measure distinct key values, BeauCoup can also be used to measure heavy hitters by estimating the number of distinct packet IDs in each flow. Specifically, BeauCoup maintains a table with bit vectors representing the coupon collectors. Upon collecting the first coupon for a flow key, BeauCoup creates a new entry in the table. When the bit vector indicates  enough coupons are collected, BeauCoup is able to tell if the threshold has been met. Since BeauCoup uses a random mapping from attributes to coupons, observing a new attribute is the same as drawing a coupon and seeing the same attribute more than once does not affect the coupon collector as it is just drawing the same coupon again. 
 


\subsection{Performance Statistics}

So far, we have discussed data structures to estimate flow-level statistics defined over a stream of individual packets and their aggregated flows. Another important line of telemetry is measuring network performance, since performance problems in the network are notoriously difficult to diagnose. Measuring performance statistics in a resource-efficient way brings new challenges to design and implement data structures for the network data plane.

Compared to measuring individual packets and aggregating the information into flows, monitoring network performance typically requires combining information across pairs of packets in a form of {\em dependency}: a packet is processed relative to some prior packet of the flow, e.g., we need to measure the time difference between a previous data packet of the flow and the corresponding acknowledgment (ACK). Given such cross-packet dependencies in the measurement, prior work~\cite{lean} has shown that it is only possible to compute performance metrics over aggregated flow statistics  (e.g., total packet loss and latency) using sublinear memory space and it is {\em infeasible} to compute other performance metrics that rely on one arbitrary pair of packets in a sublinear way, including maximum latency and maximum sending/receiving windows in the flows. The performance metrics on the flows that can be measured in sublinear memory can be called ``flow-additive''. This limitation on certain performance metrics motivates the need to design new non-sublinear algorithms that consider the problems of how to store the information about the previous packet and how to avoid bias in the estimation (e.g., bias against the traffic with larger inter-arrival or round-trip latency).




\smallskip\noindent\textbf{Round-trip time (RTT)} is a key indicator of network performance and can often be measured by the difference between the transmission time of the request packet and the receiving time of the corresponding response in a connection/flow.
Many latency-sensitive network applications, such as online gaming or trading, demand fast responses to information about new events, and are therefore extremely sensitive to latency. Hence, minimizing network latency is expected of any adequate network management. To this end, recent efforts~\cite{continuous_rtt,lean} have focused on tracking round-trip time in the data plane because traditional end host-based active probing techniques do not capture application-level RTTs (e.g., sending/receiving new packets isolated from the application) and TCP handshake-based passive monitoring can be inaccurate for long-lived connections. 

When measuring RTTs using approximate data structures, such as simple hash tables and sketches, we need to record a request in a flow first and wait for a corresponding response. However, not all requests eventually receive a response, and new requests would suffer from hash collisions with existing requests upon insertion. Upon these collisions, there are two undesirable options: (1) We can discard the new request and keep the existing one, but this will lead to a lot of stale requests staying in the data structure without a response. (2) We can overwrite the existing request with the new one, but requests from flows with larger delays may not survive for a sufficiently long time without collisions, resulting in flows with large delays being undersampled and ``bias'' towards small-delay flows. Recent work~\cite{zheng2022unbiased} proposed a data structure to correct for this measurement bias in the data plane. Specifically, they track the number of insertions into the data structure for each flow waiting for a response, and they compensate the undersampled flows accordingly when updating the data structure.

\smallskip\noindent\textbf{Out of order packets} are an indicator of the network condition and are often used to infer incorrectly configured Quality of Service (QoS). While there are quite a few root causes of out-of-order packets, faulty QoS configurations, such as setting duplicate Access Control List (ACL) rules that misclassify the packets from the same application into a different application, potentially delay some packets and fail to keep the packets in order. Such incorrect configurations can affect the performance of online applications. Thus,
network flows with many out-of-order packets can be an
indicator of current misconfigurations.

Lean algorithms~\cite{lean}, have focused on detecting flows with a high number of out-of-order packets to troubleshoot network configurations using sketches. In this work, the authors provide a definition of characterizing out-of-order packets: Given a stream of packets in one direction, the out-of-order packets are the packets whose sequence numbers are less than the current largest sequence number ($MaxSeq$), i.e., $seq < MaxSeq$, but arrive within a small period of time (e.g., 3ms) after the packet with $MaxSeq$ is received. Then their objective is to return the $k$ flows with the most out-of-order packets. To track
all flows with a high number of out-of-order packets, one needs
to compare each incoming packet against the maximum
sequence number and latest timestamp of the flow it belongs
to. Without knowing this per-flow information, a specific
packet cannot be classified as out-of-order and thus the algorithm requires per-flow memory space (non-sublinear). However, Liu et al.~\cite{lean} obtain a sublinear sketch algorithm to estimate the out-of-order packets per flow with an additional assumption: all out-of-order packets arrive within some bounded time, such as 3ms. If we do not track per-flow out-of-order packets and do not make assumptions about the bounded arrival time, recent work by Zheng et al.~\cite{zheng2022detecting} shows a sublinear algorithm to track IP prefixes that have high numbers of reordered packets and yields a better memory-accuracy bound than the algorithm in~\cite{lean}.

\smallskip\noindent\textbf{Time-to-live (TTL)} is the amount of ``hops'' or time in which a packet is set to exist inside a network before being discarded by a switch/router. In the DNS protocol, the TTL value usually specifies how long the corresponding response to a DNS record of a domain should be cached (e.g., 1-3 days).
TTL-based features in DNS protocols are particularly useful for finding malicious domains or services. As described in EXPOSURE~\cite{bilge2011exposure} and Chimera~\cite{borders2012chimera}, a number of TTL statistics from DNS responses, such as average TTL, standard deviation of TTL, number of distinct TTL values, and number of TTL changes, are indicators of malicious behaviors. This is because malicious domains and networks often have a sophisticated underlying network infrastructure.
\red{ This infrastructure can be used to change the routing path by performing attacks such as BGP hijacking which would normally} 
exhibit TTL changes. 
Additionally, it is known that a proxy running on a home network would be less reliable (usually with lower TTL values) than a server proxy running on a university environment (with higher TTL values).  
Given that these TTL-based features can be efficiently measured using sketches described above, sketch-based designs are useful to offer TTL-based performance and security monitoring capabilities.

\section{Data-Plane Resource Allocation}
\label{sec:practical}
Deploying compact data structures in network devices requires making efficient use of the limited resources in the data plane to maximize the accuracy of the query results. In this section, we explore ways to quantify measurement accuracy as a function of the size of a compact data structure, and how to optimize data-plane resources to support multiple queries. When the data plane does not have sufficient resources, queries may be partitioned to run partially in the data plane and partially in the control-plane software.  The limited data-plane resources are especially challenging to manage for long-running queries, where stale measurement data needs to expire or decay over time. 

\subsection{Quantifying Measurement Accuracy}
The choice of a particular data structure depends on the expected  accuracy.  Many network applications can tolerate a small number of false positives or slight overestimates of traffic counts.  For example, consider a firewall designed to drop unsolicited traffic entering an enterprise.  The firewall should admit incoming traffic sent in response to a recent request from an internal host, while dropping other incoming traffic.  In this setting, a Bloom filter is an appropriate data structure for storing the set of acceptable flow identifiers (based on the outgoing traffic), because the ``no false negatives'' property of Bloom filters would ensure that the firewall never drops legitimate response traffic, even if some unsolicited traffic gets through.  As another example, consider an application that probabilistically drops packets from large flows to ensure smaller flows get sufficient bandwidth.  Errors in estimating the flow counts might lead to slightly higher (or lower) drop rates for some flows, but small errors may be acceptable.

Still, the \emph{amount} of error matters.  In the firewall example, admitting a small fraction of unsolicited traffic may be acceptable, but large volumes of unwanted traffic would defeat the purpose of having a firewall.
Past theoretical work has led to analytical error bounds for a number of compact data structures~\cite{ams,cms,CountSketch}.  Yet,  many of these data structures require modification to work in the data plane---particularly to handle limitations on memory-access bandwidth.  Analysis of compact data structures designed for the data plane is an exciting avenue for future research.  However, analytical results are often quite conservative.  Accuracy is often much higher under realistic traffic than the models would suggest. For example, network traffic often follows a Zipfian distribution, with a small number of large flows and a large number of small flows.  Traffic is often bursty, with packets of the same flow arriving close together in time.  Incorporating assumptions about the traffic distribution into the analysis could lead to analytical models that provide better estimates of measurement accuracy.

Rather than relying on analytical results, simulations on representative traffic traces can drive the choice of data-structure parameters.
Network operators can collect packet traces from their own networks, and use these traces to simulate a candidate data structure and compare the measurement results with ground truth.  Using local packet traces has the advantage of capturing a realistic workload for the network in question, though network operators may not know how long of a trace to use or how diurnal patterns or other traffic shifts might affect the suitability of the data structure.  Researchers often use packet traces, too, including publicly available measurement data to better understand how compact data structures perform for realistic traffic across a range of data-structure parameters (e.g., the number of rows $r$ and columns $c$ in a count-min sketch).

An alternative approach is to have the data structures provide estimates of their own measurement error at run time.  Run-time reporting of measurement error has the advantage of automatically reacting to changes in the traffic distributions, and providing actionable information to network administrators or the network itself. For example, high measurement error could make a network administrator more conservative in deciding to block or rate limit seemingly suspicious traffic, or could lead to changes in the choice of data structure. Some compact data structures naturally provide a way to quantify their own accuracy. For example, the current values of the $r*c$ counters in a count-min sketch can be used to compute tighter bounds on the estimation error for traffic counts~\cite{error-estimate}. Designing new kinds of ``self-measuring'' data structures is an exciting avenue for future research.

\subsection{Optimizing Resource Allocation}
Deploying compact data structures in practice relies on making decisions about the final size and shape of the structures.  Fortunately, approximate data structures are, by their nature, \emph{elastic}; that is, they remain valid under a variety of configurations.
For example, for a count-min sketch, increasing the number of rows $r$ or columns $c$ (or both!) increases accuracy, but at the expense of consuming more of the limited data-plane memory. 
Given resource constraints, a compiler can determine the values of $r$ and $c$ that, together, maximize measurement accuracy, subject to ``fitting" within the data-plane target~\cite{p4all,sun2024autosketch}. When the data plane must support multiple telemetry tasks, the compiler can select parameters for each data structure to maximize some weighted objective function that considers the accuracy of each query, subject to all of the data structures fitting within the resource constraints. 

Rather than statically allocating a separate data structure for each query, the data plane could use a single shared data structure. This approach has the  \emph{statistical multiplexing} advantage of the limited data-plane memory. However, a shared data structure imposes more limits on  the number of queries each arriving packet can update.  In the simplest case, different queries operate on different traffic, allowing the telemetry system to associate each packet with only one query. More generally, though, queries consider overlapping traffic. For example, one query may identify heavy-hitter source IP addresses, while another identifies heavy-hitter destination IP addresses, and yet another identifies destinations receiving traffic from a large number of distinct sources.  Each packet is relevant to all three of these queries.  In this setting, sampling can enable each packet to update state for at most one of these three queries, to stay within the data plane's limited memory access bandwidth~\cite{chen2020beaucoup}. \red{Furthermore, Sketchovsky~\cite{sketchovsky} showed that creating an ensemble of sketches that can allow reuse of hash results in addition to improving memory usage.}

\red{Recent work explores} a division of time across queries. In one solution, each query is allotted a certain interval of time in which it is performed, according to both the available resources and the accuracy required for the query~\cite{dynatos}. In other solutions, queries can be modified on-the-fly, during run-time, enabling the user to actively decide which queries should be performed, according to the changing needs of the network\red{~\cite{flymon, chamelemon,sun2024autosketch,laffranchini2019measurements, queryPlanningConext24,srivastava2024raising}}.

\subsection{Partitioning the Queries} 
Unfortunately, the data-plane resources may not be sufficient for computing accurate answers for each of the queries.  In 
some cases, the memory size or bandwidth may be too small to handle every query. In other cases, queries may require complex operations---such as regular expressions on packet payloads, or computing the median of a large set of numbers---that the data plane cannot perform.  Rather than sacrificing accuracy or functionality, a telemetry system can \emph{partition} a query to run partially in the data plane and partially in the control plane~\cite{sonata}. For example, a class of Telnet-based attacks on Internet of Things (IoT) devices involves the adversary sending short Telnet packets with a particular command (e.g., ``ZORRO'') in the payload~\cite{zorro}. While parsing strings in the packet payload may be too expensive, 
the data plane can readily identify small packets using the Telnet transport port, and direct them to control-plane software for further analysis of the packet payload~\cite{sonata}

Alternatively, the telemetry system can capitalize on extra storage space outside of the data plane, to help in accumulating statistics that cannot fit entirely in the data-plane registers~\cite{marple}. For example, the data plane could store and accumulate statistics for the popular active flows, and maintain information about other flows in a slower
memory in the control plane.  This approach works well for certain statistics, such as the total number of bytes or packets in a flow.  However, merging the aggregate statistics from two locations is not always possible without some loss in accuracy, depending on the statistic of interest. For example, consider a query that counts the number of out-of-order packets in a TCP connection. One way of defining this query is to count the number of packets  with sequence numbers smaller than the maximum value seen so far. However, the maximum sequence number may appear very early in a long stream, making it difficult to merge results from the data plane with earlier results stored in the control plane~\cite{marple}.

In these kinds of telemetry systems, a compiler needs to \emph{partition} the set of queries to determine what portion of the analysis can run in the data plane, and what portion must rely on separate processing and memory resources, while minimizing the overhead of involving the control plane.
Another way to involve the control-plane software is in 
analyzing samples or measurements gathered in the data plane.
Some recent telemetry systems~\cite{aroma,FlowRadar,univmon}
collect flow samples or flow counters within the programmable data plane and perform further analysis in the control plane to reconstruct the statistics of interest.
Despite enabling a richer set of queries, dividing the analysis between the data-plane hardware and the control-plane software often comes with a price. First, the split may incur significant communication overhead. Second, analysis will take orders of magnitude more time than analysis that is performed directly and entirely in the data plane. 

\subsection{Reusing Data Structures Over Time}

\cut{Since measurement data becomes less relevant with time, telemetry systems typically produce statistics about the most recently seen traffic.  For example, a network operator may want to know the top-ten flows over a $30$-second period.
With a \emph{tumbling} window, the time intervals do not overlap, and each packet is processed once and belongs to one window. For example, the first interval would correspond to times $0$ to $30$ seconds, while the next corresponds to times $30$ to $60$ seconds.
In contrast, a \emph{sliding} window slides over the stream of packets, always maintaining statistics for the most recent packets.  Sliding windows are more expensive to maintain, since the older packets expire gradually. In practice, a sliding window may be approximated using multiple smaller tumbling windows (e.g., times $0$-$10$, $10$-$20$, and $20$-$30$ for the $0$-$30$ second interval, followed by times $10$-$20$, $20$-$30$, and $30$-$40$ for the next interval covering times $10$-$40$).\jrex{move to section 2?}}


The data plane cannot maintain traffic statistics for long intervals of time, since, as the structures fill up, accuracy may be reduced due to excessive collisions. Furthermore, counters could overflow, causing significant errors in counter values. 

One way to clear  the structure is by decaying old statistics, using methods such as exponential weighted moving average \red{(EWMA). While it is difficult to perform exact EWMA in the data plane, recent work performs approximate EWMA~\cite{hhIPG}. } 
Another approach is to collect new traffic statistics as time passes by evicting old data~\cite{recento}, or by reinitializing parts of the data structure while continuing to collect new data.  
For example, the data plane could maintain four data structures, each covering a ten-second period, which will be used as a sliding window. One data structure can be cleaned (i.e., its counters reset to $0$) while the other three are used to collect statistics for a $30$-second interval. For example, the data structure for times $0$-$10$ could be cleaned during times $30$-$40$, and then populated again during times $40$-$50$ to contribute to the statistics for the intervals $20$-$50$, $30$-$60$, and $40$-$70$, respectively. 
\red{Sliding windows allow clearing out the data in bulk, while decaying counters using EWMA, for example, allows greater granularity. However, it is harder to implement in the data plane. 
While a recent work presents a Bloom filter that supports deletion and decay~\cite{setd4},
the topic of decay has not been studied much within the restrictions of the data plane, and remains a promising direction for future work.}

Alternatively, instead of using a decaying or sliding window mechanism, a recent solution proposes a new paradigm of \emph{monitoring on demand}~\cite{MIDST}. 
There, the system measures queue-based loss. Such loss can only occur when there is build-up in the queue, and will not happen otherwise. Therefore, the system starts monitoring when required, that is when the queue starts to build up, 
and stops when the queue winds down. At that point, the structure is cleaned and prepared for the next monitoring interval.


\textbf{Cleaning data structures. } The responsibility for cleaning a data structure can fall to separate control-plane software that accesses the data-plane registers to reinitialize the values.  However, the cleaning operations can introduce control-plane overhead and the need for close time synchronization between the control and data planes.  An alternative is to ``clean'' the stale data structure directly from the data plane, as the packets fly by~\cite{conquest}.  Each packet, then, would trigger the data plane to clean a portion of one data structure, while reading from the remaining structures and writing to the most recent structure.  

During periods of low traffic loads, the data plane may not receive sufficient traffic to complete the cleaning process before the data structure must return to collecting statistics.  The switch can ensure sufficient ``cleaning traffic'' by recirculating a small amount of traffic simply to trigger the data plane to reset the elements of the stale data structure.
Another technique to speed up cleaning is based on maintaining a smaller representation of the data structure, where each register is represented by a single bit \red{(i.e., a bitmap)}. 
Instead of cleaning the entire data structure, the representation is reset, requiring significantly fewer packets for the process~\cite{MIDST}. Then, for each register, the next time it is written to, if its representative bit is still set to $0$, the register will first be cleaned, and the bit set to $1$, to avoid unnecessarily resetting the register while it is already in use.





\section{Distributed Telemetry} 
\label{sec:distributed}
\cut{
\alan{more related work digging up}
\shir{Focus on open problems. 7.1 telemetry across ports in a switch - we want the abstraction of a single structure across multiple pipelines, 7.2 telemetry across switches in a network - one-big-switch, 7.3 decentralized collection of data - what did a packet experience across its journey (e.g., INT) vs. aggregate or sub-select traffic along a path (which flows have a certain amount of packets that experienced long queuing delays for example). INT does not have sketches but if we analyze the data in the data-plane sketches could be useful.  }

(which introduces communication cost and decisions about placement of function), presented more as specific case studies
\jrex{each subsection has: (i) motivation for telemetry in this setting, (ii) what makes it challenging, and (iii) initial results so far.  we can cite some "non-compact" work to illustrate the motivation and what is known so far.}
}
In the prior sections we have discussed ways to measure traffic at a single location in the network. However, network administrators often need to collect measurements from multiple vantage points and combine or merge them to get a broader view of the state of the network. 
In this section we discuss the challenges of collecting and analyzing measurements from across the network and specifically the need for efficient coordination.



Coordination poses an inherent trade-off; on the one hand, the amount of information collected at each location, as well as the frequency and size of the coordination messages, need to be limited, yet harsh limitations may degrade the accuracy of the system. We thus need to find ways to optimize this communication.  
We discuss two main paradigms for performing coordination for distributed telemetry. The first piggybacks information on packets traversing the network; that is, packets ferry state about their own experience or the state of the network.  The second makes use of dedicated coordination messages. These paradigms can be used separately or combined, depending on the task at hand.

In this section, we will show how each of these paradigms can be useful for different network tasks and survey existing solutions.
Despite promising advances in these areas, the topic remains mostly uncharted, leaving many problems open for further research. 


We note that an underlying assumption in our discussion is that measurements may be combined. 
Looking at basic data structures that we saw in Section~\ref{sec:heavyhitters}, 
certain sketches such as the count-min sketch, are known to be mergeable (assuming they are of the same size and functionality). However, merging other types of data structures (e.g., Space Saving) may be non-trivial. The need for network-wide measurements 
requires us to find ways to merge these data structures in an efficient manner, yet this will not be the focus of our discussion.

\subsection{Network-wide Measurements}

In Section~\ref{sec:heavyhitters}, we saw data structures for identifying heavy hitters in a single device. Yet, network problems often spread throughout the network.
For example, often in DDoS attacks, the attack traffic may not be heavy at a single point, but the sum of the traffic is substantial and constitutes a \emph{network-wide heavy hitter}.
In order to detect the attack in this scenario, measurements performed in various locations in the network need to be aggregated.  
The aggregation of measurements performed across the network provides the abstraction that the network is a single measurement entity, also known as the \emph{one-big-switch} network abstraction.



One of the key challenges in network-wide monitoring is task placement, that is where in the network should  measurements be performed. Systems such as vCRIB~\cite{vcrib} provide a solution for source partitioning in host-based network measurements in data centers, that is both resource-aware and traffic-aware. With the rise of network functions and the wide range of programmable devices, the options are vast and there is much opportunity for optimizing task placement in the network~\cite{networkTaskPlacementCCR20, HeteroSketch,chen2024eagle}, as well as more advanced coordination between the measurement points~\cite{Mozart}. \red{In addition to placing tasks among switches or hosts, some systems coordinate a network-wide deployment among hosts, switches, and controllers~\cite{omniwindow,huang2020omnimon}.}




Another challenge is to determine the rate at which the aggregation is performed. Most existing network-wide telemetry solutions perform measurements (or collect samples) in various locations, which are then reported back to a controller for aggregation~\cite{elasticsketch,univmon}. 
The reporting rate therefore has a significant impact on the accuracy of the aggregated measurements and while higher reporting rates will usually lead to higher accuracy, they may also incur high communication overhead. Many solutions use a fixed reporting rate~\cite{aroma, Harrison_network_wide_hh, networkWideHHIncrementalDeployment}. An alternate approach seeks to perform continuous monitoring~\cite{ContinuousDistMonitoring,zhang2024octosketch}, using probabilistic reporting~\cite{carpe}. 



\subsection{Telemetry Across a Path} 


Monitoring the experience of a packet traversing the network is useful for pinpointing problems and improving performance. Information can be collected about the path that the packet traversed and what the packet experienced along this path. 
For example, as a packet goes through a switch, it can detect the delay it experienced in the queue by computing the difference between the time it entered the queue and the time it left the queue. As the packet traverses the network, it can aggregate the queuing delay it experienced across the entire path. 
Path-based queries can include any statistic relevant to a packet, flow or network devices, including queuing loss and flow counts~\cite{compilingPathQueries, LightGuardian} or congestion~\cite{PassiveIPMeasurements}. 
Path based monitoring may even be used for detecting the source of spoofed packets~\cite{IPTraceback, HashBasedIPTraceback}.  
\red{Data carried by the packets can also be accumulated inside the data plane by the switches along the path of the packet. For example, structures for finding heavy hitters can be used to identify the paths or flows experiencing heavier delays.}

The inband network telemetry (INT) framework, as found in certain network devices~\cite{deepInsight}, enables telemetry to be collected and$/$or aggregated within the data plane without the need for controller involvement~\cite{INT, intDemo}. In the INT framework, packets may carry measurements and data as well as telemetry instructions that are read and executed by the network devices. The collected measurements are usually sent back to the controller for further analysis, yet it is possible to perform some of the analysis inside the data plane for in-network computation. For example, information on queuing delay can be accumulated in a structure for heavy hitter detection to identify flows that were heavily delayed.

Programmable networks provide the 
flexibility to collect statistics as defined by the programmer. 
Yet, if 
 many INT tasks are being performed, a lot of data may be appended to each packet, thus increasing the load on the network. 
Yet, programmable devices also enable aggregating the statistics rather than maintaining per-hop information along the entire path. 
Furthermore, solutions such as PINT~\cite{pint}, provide a probabilistic variation of INT that bounds the per-packet overhead while providing similar monitoring capabilities. 


\begin{figure}
\centering
\includegraphics[trim = 0mm 40mm 25mm 0mm, clip,scale=0.55]{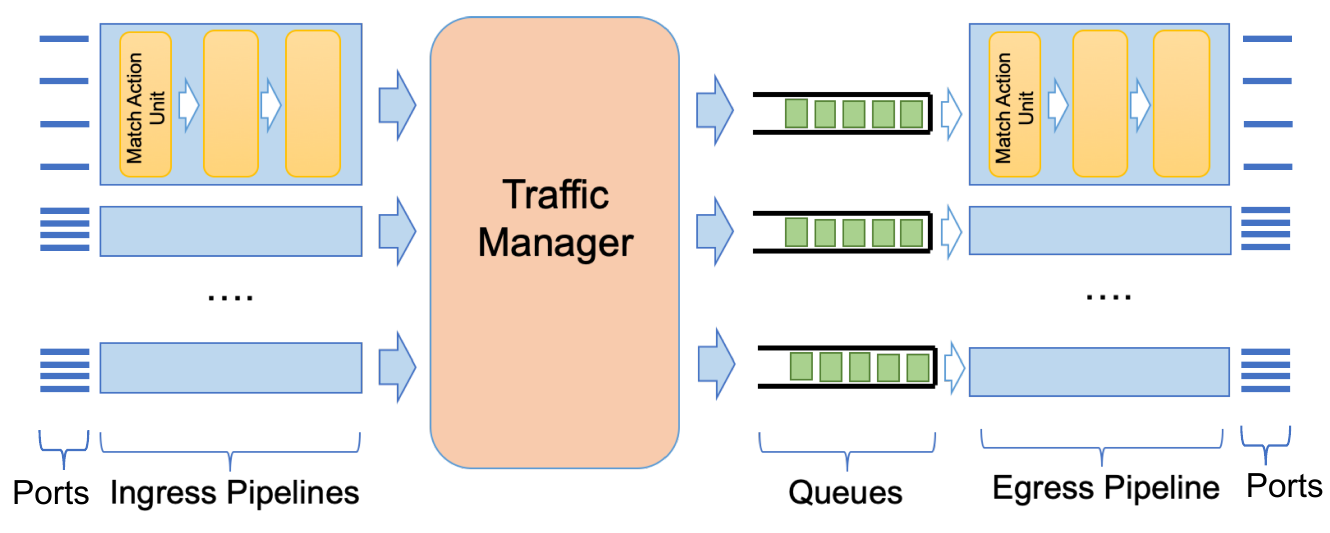}
\caption{PISA Programmable Switch Architecture \red{with  multiple ingress and egress pipelines. Each port is coupled to a dedicated pipeline. Each packet comes in through an ingress port and traverses the ingress pipeline coupled to this port. The egress port will be determined by the ingress pipeline. The traffic manager receives the packet from the ingress pipeline and forwards it to the queue of the egress pipeline that is coupled to this egress port.}}
\label{fig:PISAModel}
\end{figure}

\subsection{Intra-Switch Distributed Telemetry}

While we often think of distributed telemetry as spanning multiple locations in the network, even collecting telemetry in a single switch is inherently distributed. 
As seen in Figure~\ref{fig:PISAModel}, programmable switches often have more than one ingress or egress pipeline~\cite{BroadcomTomahawk4}. 
Each pipeline serves a given set of ports. Upon entering the switch at an ingress port, a packet is processed by the ingress pipeline that determines which egress port, and respective egress pipeline, should handle the packet as it leaves  the switch. 
Thus, each ingress pipeline potentially feeds all of the egress pipelines. 


We consider two main types of intra-switch telemetry.
The first measures intra-switch statistics such as loss or delay. 
Such information may need to be collected across multiple disjoint pipelines within a single switch.
While many solutions perform measurements individually for each pipeline~\cite{precisionToN, MIDST}, in the case of multi-path routing or load-balancing, flows may arrive at different ingress pipelines or exit through different egress pipelines. In this case, collecting any flow-based measurements requires aggregating the counters found in the different pipelines of the switch.

The second type performs joins across pipelines. For example, we might wish to match between a SYN and the respective ACK~\cite{liu2021jaqen} (or compute the RTT~\cite{continuous_rtt, RTTSPIN20} of a flow). Even if both the SYN and ACK packets traverse the same path in the network, they could still be processed by different ingress and/or egress pipelines. Thus, the match or \emph{join} would need to be performed across the different pipelines. 

Due to the compartmentalized memory model in programmable switches~\cite{IntelP416}, integrating measurements across pipelines is quite challenging.  
Certain solutions attempt to overcome this obstacle by ferrying information between ingress and egress on existing traffic~\cite{HHMultiPipeline,MIDST} 
or by transferring information between pipelines using designated packets~\cite{swish}. In these solutions, information passed between pipelines is used to merge or aggregate measurements across pipelines to provide the abstraction of a single structure for all pipelines. Other solutions divide the flow space between pipelines to avoid the need for aggregation~\cite{MultiPipeChiesa}.

Recently, we have also seen somewhat orthogonal solutions that suggest an extension to the existing data-plane architecture in order to support stateful packet processing across pipelines~\cite{stateLessQuo, MultiPipelineSIGCOMM22}.

\section{Security}
\label{sec:security}

In recent years, programmable networks have been used to enhance network security~\cite{liu2021jaqen, zhangposeidon, PigasusSherry}, yet apart from a handful of works~\cite{P4Vulnerabilities, P4SecurityVulnerabilitySurvey}, the \emph{vulnerability} of programmable networks and devices has received less focus. In this section, we explore some of the more common vulnerabilities of programmable devices with respect to network telemetry.  

\subsection{Better Hash Functions}

One of the key components of most measurement data structures in the data plane is the \emph{hash function}. To maintain the processing at line rate, current hardware performs hash functions using hardware implementations. One such common implementation is based on the cyclic redundancy check (CRC). The fixed-length output of the CRC was originally intended to be an error detection code, and is being used in many applications as a hash output. However, CRC is not a cryptographic hash function.  The short hash length is vulnerable to collision-based attacks and is therefore considered insecure~\cite{wang-adversary}.
Recently, we have seen a solution for implementing a variant of SipHash~\cite{siphash} in the data plane~\cite{SipID}, yet the process requires multiple pipeline passes and thus does not keep up with line rate.  
We have also seen a solution which implements the cryptographic function AES~\cite{AESDannySPIN} on a programmable switch, yet this solution too requires multiple recirculations to be performed. A faster alternative has been presented using the Even-Mansour scheme that is able to perform packet-header obfuscation at switch hardware rates~\cite{pinot}. These results show that this is a promising direction, yet more robust solutions are still needed.



\subsection{Robustness to Adversaries}
The data structures themselves can also be vulnerable to attack and adversarial traffic. For example, adversarial traffic may pollute Bloom filters causing a high collision rate and consequently an increase in false positive rate~\cite{SketchesAdversarial}, or cause count-min sketches to incorrectly detect large flows~\cite{CMSAdversarial}.
Certain adaptations to these structures have been presented. First, to be more resilient to pollution, the sketches should be as large as possible. For example, the Broom filter~\cite{broom} divides the filter between local and remote memory, so the overall size of the structure can be much larger. The overhead caused by access to remote memory is reduced due to a unique feedback mechanism between remote and local memory. 
Second, the standard hardware hash functions are publicly known. Thus, randomness (i.e. \red{a random value or} salt) or a secret key should be used when computing hash functions, to prevent an adversary from determining the output of the hash functions and using this information to modify the data~\cite{SketchesAdversarial}. Third, network devices may be vulnerable to various software and hardware bugs, which can allow attackers to gain unauthorized access to sketch data structures and modify the output telemetry results. TrustSketch~\cite{trustsketch} proposes and optimizes running sketches inside trusted execution environments (TEEs) such as Intel SGX to preserve the integrity of the sketch data structures and the telemetry output. 

To enhance the security of network telemetry, and specifically sketch-based measurements, we believe that finding additional vulnerabilities in the structures as well as devising \emph{practical} ways to secure them is a problem that should be further explored.

\section{Conclusions}
\label{sec:conclude}
The emergence of high-speed programmable network devices offers the potential for unprecedented visibility into network traffic.  Compact data structures are critical for bridging the gap between the questions network administrators want to answer and the limited computation and memory resources on network devices.  The research on compact data structures for network telemetry over the last several years gives a sense of the exciting possibilities, yet so many important research challenges remain.  Future research can consider more sophisticated queries (including network performance), as well as better ways to support multiple queries concurrently and to change the collection of queries at run-time. Distributed telemetry and robustness to adversaries are other key directions for further work.  Finally, future work can explore \red{how the network devices may take action automatically, to close the control loop.  
This will allow telemetry systems} to go beyond analyzing the traffic to taking actions on traffic directly in the data plane to improve network performance, security, and reliability. 

\section*{Acknowledgments}
This work is supported in part by the Israel Science Foundation (grant No. 980/21)
and the U.S. National Science Foundation (grant CNS-1704077). Liu is supported in part by the U.S. National Science Foundation CNS-2431093.

\clearpage
\bibliographystyle{acm}
\bibliography{references}

\end{document}